\documentclass[aps,pra,twocolumn,superscriptaddress]{revtex4}
\usepackage{amssymb}
\usepackage{graphicx}
\usepackage{amsmath}
\usepackage{verbatim}

\begin{document}

\title{Sigature of the universal super Efimov Effect: three-body contact in two dimensional Fermi gases}
\author{Pengfei Zhang}
\affiliation{Institute for Advanced Study, Tsinghua University, Beijing 100084, China}
\author{Zhenhua Yu}
\email{huazhenyu2000@gmail.com}
\affiliation{School of Physics and Astronomy, Sun Yat-Sen University, Zhuhai 519082, China}

\date{\today }

\begin{abstract}
A new class of universal ``three-body" bound states has been recently predicted theoretically for identical fermions interacting at $p$-wave resonance in two dimensions. This phenomenon is called the super Efimov effect since the binding energies of the states follow a intriguing double exponential scaling. However, experimental resolution of this scaling is expected to meet formidable challenges. In this work, we introduce a new thermodynamic quantity, the three-body contact $C_\theta$, to quantify three-body correlations in a two dimensional gas composed of the resonantly interacting fermions; the contact $C_\theta$ is the consequence of the underlying universal super Efimov effect in the many-body context. We show how $C_\theta$ affects physical observables such as the radio-frequency spectrum, the momentum distribution and the atom loss rate. Signature of the elusive super Efimov effect in the thermodynamic system can be pinned down by the detection of the three-body contact $C_\theta$ via these observables.  
\end{abstract}

\maketitle

Correlations lie at the heart of interacting systems. Crucial properties of unitary atomic Fermi gases, despite its nature of strong interactions, 
have been understood both theoretically and experimentally thanks to the introduction of the concept -- ``contact" \cite{Tan2008, Braaten2008, Zhang2009, Werner2009, Stewart2010, Sagi2012, Hoinka2013}. The successful application of ``contact" is based on the insightful observation that in dilute atomic gases, usually, the probability of finding three or more atoms at short distances is so rare that the two-body correlations become the primary; the contact quantifies the strength of such two-body correlations. Theoretical generalization of contact for $p$-wave interacting gases \cite{Ueda2015, Yu2015, Qi2016, Ueda2016, Qi2016x} managed to explain the distinct two-body correlation structures discovered experimentally in $^{40}K$ gases close to $p$-wave resonances \cite{Thywissen2016}. Extension of the concept to other situations has also been considered \cite{Valiente2011, Valiente2012, Yu2016, Cui2016, Qi2016r, Yi2016x}.

On the other hand, the Efimov effect is known to give rise to observable three-body correlations in three dimensional gases \cite{Braaten2011, Jin2014, Braaten2014, Hadzibabic2016}. In 1970, V.~Efimov showed that when three bosons interact at $s$-wave resonance in three dimensions, there exist a series of bound states whose binding energies $E_n$ obey a geometric scaling $E_n=E_*e^{-2\pi n/s_0}$, with $s_0\approx 1.0062$ a universal constant \cite{Efimov1970}. The Efimov bound states are understood to originate from an effective potential $U_{\rm E}(\rho)=-(s_0^2+1/4)/\rho^2$ with $\rho$ the hyper-radius, when the three-boson problem is formulated in the hyper-spherical formalism. An extra three-body parameter $E_*$ is needed to regulate the singularity of the effective potential at small $\rho$. At the few-body level, the measurement of atom loss due to three-body recombinations has demonstrated the geometric scaling of the universal Efimov effect \cite{Grimm2006,  hulet, Gross1, gross, Berninger}. 
In a thermodynamic many-body system, E.~Braaten \emph{et. al.} pointed out that the Efimov effect shall give rise to a ``three-body" contact $C_3$ which relates to the change of the system energy $E$ as $C_3\sim \partial E/\partial E_*$ \cite{Braaten2011}. This thermodynamic quantity $C_3$ measures the probability of three atoms being close to each other, and takes part in various universal relations. The subleading contribution of the momentum distribution tail detected in the dynamic unitary Bose gas experiment \cite{Jin2014} is attributed to the three-body contact $C_3$ \cite{Braaten2014}. A recent Ramsey interferometry experiment has nailed down non-zero $C_3$ in the coherent evolution of resonantly interacting Bose gases as well \cite{Hadzibabic2016}.

In two dimensions, Y.~Nishida \emph{et. al.} discovered a new class of universal three-body bound states for identical fermions interacting at a $p$-wave resonance \cite{Nishida}. These states are called super Efimov states as their binding energies follow an intriguing double exponential scaling $E_n\sim \exp({-2e^{3\pi n/4+\theta}})$. Subsequent hyper-spherical formalism calculations revealed that the super Efimov bound states derive from an effective one-dimensional eigen-equation for the reduced three-body wave-function
\begin{align}
\left[-\frac{d^2}{d\rho^2}+U_{\rm SE}(\rho)\right]f=Ef,\label{hyper}
\end{align}
with the effective potential $U_{\rm SE}(\rho)=-1/(4\rho^2)-(16/9+1/4)/(\rho\ln\rho)^2$ \cite{Gao2015, Zinner}. According to Eq.~(\ref{hyper}), in the asymptotic region $\rho\sqrt{E}\ll1$, the three-body wave-functions of the super Efimov states are superpositions of the incoming solution $\sqrt{\ln\rho}\exp[-i(4/3)\ln(\ln\rho)]/\rho$ and the outgoing one $\sqrt{\ln\rho}\exp[i(4/3)\ln(\ln\rho)]/\rho$; the ratio of the superposition coefficients is $\pm e^{i8\theta/3}$, fixed by the dimensionless three-body parameter $\theta$.
Calculations employing various model potentials with a van der Waals tail indicate that while the overall scale of the binding energies is associated with the van der Waals length $\ell_{\rm vdW}$, the three-body parameter is $\theta\approx-1.5$; the ground super Efimov state is predicted to emerge at the threshold where the two-dimensional $p$-wave scattering area is about $-42.0\,\ell_\text{vdW}^2$ \cite{Gao2015}. Atom loss enhancement expected at the emergence can serve as a signal of the super Efimov effect. However, while resolving the double exponential scaling is still possible in dynamic expansions where relevant scales can be engineered appropriately \cite{Shi2016, Yu2016ex}, experimental detection of the same super Efimov effect in atomic samples is formidably challenging. 

In this work, we introduce the three-body contact $C_\theta$ for a two dimensional gas of identical fermions close to a $p$-wave interaction resonance. This new thermodynamic quantity, defined as the derivative of the system energy 
\begin{align}
C_\theta\equiv&\frac{\partial E}{\partial \theta}, \label{ctheta}
\end{align} 
is the consequence of the underlying super Efimov effect in the many-body system. We show how $C_\theta$ affects universal relations of the system, regarding the radio-frequency spectrum, the momentum distribution, and the atom loss rate. Our results lay out a roadmap for measuring the three-body contact $C_\theta$ via these relations, and thereby provide an alternative opportunity to detect the signature of the elusive super Efimov effect in the thermodynamic system.

Since the early studies of degenerate atomic gases, the radio-frequency spectroscopy has been an widely used experimental method to diagnose interaction effects \cite{Kleppner1998, Pethick2001, Gupta2003, Zwierlein2003, Chin2004, Yu2006, Baym2007, Jin2008, Zwierlein2009, Hu2010}. By the Fermi golden rule, the radio-frequency spectrum $I_{\rm rf}(\omega)$ is given by
\begin{align}
I_{\rm rf}(\omega)=&-\dfrac{1}{\pi}{\rm Im}\Pi_{\rm rf}(\omega), \\
\Pi_{\rm rf}(\omega)=&-i\int dt \,e^{i\omega t}\langle T \mathcal O(t)\mathcal O^{\dagger}(0)\rangle,
\end{align}
with $\mathcal O(t)\equiv\int d^2\mathbf{r} \psi^{\dagger}_{e}(\mathbf{r},t)\psi(\mathbf{r},t)$, $\psi$ the fermion field operators and $T$ the time ordering operator. The operator $\mathcal O$ accomplishes the action of the external radio-frequency field: changing the internal state of a fermion into an excited one labelled as $e$.

The super Efimov effect gives rise to a large frequency tail of the spectrum
\begin{align}
I_{\rm rf}(\omega)_{\rm 3B}=\frac{C_\theta}{3\omega^2\ln(\sqrt{\omega}R)}f(\omega)
\label{i3}
\end{align}
with 
\begin{align}
f(\omega)=74\cos[2\Theta(\omega)]-20\sin[2\Theta(\omega)]-83
\end{align}
and $\Theta(\omega)=\frac{4}{3}(\ln[-\ln(\sqrt{\omega}R)]-\theta)$. Here $R$ is the effective range of the pairwise $p$-wave interaction. The presence of $R$ is because one can parameterize the $p$-wave scattering phase shift in the low energy limit as $\cot\delta(k)=-1/(sk)+(2/\pi)\ln(kR)$. Close to a resonance, the scattering area $s$ is divergent and $R$ remains finite; the effective range $R$ becomes a natural scale to distinguish short and long wavelengths.  

Similar as the unitary Bose gases in which the Efimov effect exists \cite{Braaten2011, Braaten2014}, two-body correlations due to pairwise interactions also contribute to the spectrum tail. Given that the two-body interaction is parameterized by $s$ and $R$, we define two ``two-body" contacts
\begin{align}
 C_s\equiv&-\frac{\partial E}{\partial s^{-1}},\label{cs}\\
 C_R\equiv&-\frac{\partial E}{\partial R^{-1}},\label{cr}
 \end{align}
which quantify the probability of and the one weighted by relative energy of finding two fermions at short distances respectively; counterparts of these two ``two-body" contacts have been introduced to $p$-wave gases in three dimensions \cite{Ueda2015, Yu2015, Qi2016}. The corresponding spectrum tail is given by
\begin{align}
I_{\rm rf}(\omega)_{\rm 2B}=&\frac{2}{\pi\omega}C_s+\frac{1}{\omega^2R}C_R.\label{rf2}
\end{align}

On the face of it, $I_{\rm rf}(\omega)_{\rm 3B}$ is eclipsed by $I_{\rm rf}(\omega)_{\rm 2B}$; the three-body contact contribution differs from the sub-leading term $\sim C_R$ in Eq.~(\ref{rf2}) by a logarithm of frequency $\omega$. However, the recent dynamic measurement of three-dimensional $p$-wave Fermi gases has managed to determine not only the leading but also the sub-leading behavior of the radio-frequency spectrum \cite{Thywissen2016}. Moreover, once the gas is confined in a two-dimensional harmonic trap, the system satisfies the modified virial theorem
\begin{align}
E+C_s/s+C_R/2R=2V_{\rm ho},\label{virial}
\end{align}
with $V_{\rm ho}$ the harmonic trap potential energy of the system. 
The relation (\ref{virial}) is based on dimensional analysis and $s$, $R$ and $\theta$ being the only relevant interaction parameters.
The absence of $C_\theta$ in Eq.~(\ref{virial}) is the consequence of $\theta$ being dimensionless, contrary to the Efimov case \cite{Braaten2011}.
At least on resonance $1/s=0$, $C_R/R$ can be solely determined via the virial theorem by measuring the potential and the total energy independently \cite{Stewart2010}. This simplification may ease experimental extraction of $C_\theta$.

Time-of-flight provides another useful tool to reveal correlation effects in the momentum distribution $n(\mathbf q)$ of atomic gases \cite{Bloch2008}. Due to the angle dependence of $p$-waves, $n(\mathbf q)$ in principle depends on the direction of $\mathbf q$ \cite{Ueda2015, Yu2015, Qi2016, Ueda2016, Qi2016x}. 
However, (partial) direction average is usually taken to enhance signals at large momentum for fermions \cite{Stewart2010, Thywissen2016}. 
The contribution from the three-body contact to the direction averaged momentum distribution $n(q)$ for large $q$ is given by
\begin{align}
n(q)_{\rm 3B}=\frac{\sqrt2 C_\theta}{\pi}
\frac{\cos[2\tilde\Theta(q)+\pi/4]}{q^4\ln(\sqrt{3}qR/2)}\label{n3},
\end{align} 
with $\tilde\Theta(q)=\frac{4}{3}(\ln[-\ln(\sqrt{3}qR/2)]-\theta)$. 

Certain features of Eq.~(\ref{n3}) reflect the behavior of the wave-functions of the super Efimov states which can be derived by the hyper-spherical formalism. Due to the effective potential $U_{\rm SE}(\rho)$, the hyper-radial parts of the super Efimov wave-functions are asymptotically the superpositions of two linearly independent solutions $\sqrt{\ln(\rho/R)}\exp[-i(4/3)\ln(\ln\rho/R)]/\rho$ and $\sqrt{\ln(\rho/R)}\exp[i(4/3)\ln(\ln\rho/R)]/\rho$ with the superposition coefficients being $\pm e^{i8\theta/3}$ \cite{Gao2015, Zinner}. When one calculates the tail of $n(q)$ from the Fourier transform of the three-body wave-functions, the $1/q^{4}$ factor in $n(q)$ comes from the $1/\rho$ factor in the spatial wave-functions, while the $\cos[2\tilde\Theta(q)+\pi/4]$ part corresponds to the oscillation of the wave-functions in the real space.

Likewise, the two-body contacts contributes to $n(q)$ as
\begin{align}
n(q)_{\rm 2B}=\frac{2C_s}{\pi^2q^2}+\frac{2C_R}{\pi R q^4}+\frac{2C_{s,P}}{\pi^2q^4}\label{t-b}.
\end{align}
There is an additional quantity $C_{s,P}$ in Eq.~(\ref{t-b}), which quantifies the probability of finding two fermions at short distances weighted by their center of mass energy (see Eq.~(\ref{csp})). The general unavoidable presence of $C_{s,P}$ in the momentum distribution can be understood by considering a single pair of interacting fermions and inspecting the transform of the momentum distribution between different reference frames \cite{Yu2016}. 
Comparison between the momentum distribution (\ref{n3}) and (\ref{t-b}) and the radio-frequency spectrum (\ref{i3}) and (\ref{rf2}) suggests a correspondence $q^2\leftrightarrow \omega$ in respective terms, though the exact prefactors can only be worked out through detailed calculations.

Enhanced three-body recombination rate observed in Cs consists of the first experimental signature of the universal Efimov effect in atomic gases \cite{Grimm2006}.
Since the three-body contact $C_\theta$ physically quantifies the probability of finding three fermions at short distances where three-body recombination processes occur mostly, the three-body atom loss rate shall relate to $C_\theta$. This rate can be calculated by introducing an imaginary part to the system Hamiltonian $H$ such that the three-body parameter $\theta$ acquires an imaginary part $-i\eta$ correspondingly \cite{Braaten2011}. We find that the total number $N$ of the fermions satisfies
\begin{align}
\frac{dN}{dt}=-i(N H-H^*N)=2N\text{Im} H.
\end{align}
When $\eta$ is small, we obtain the three-body loss rate
\begin{align}
\Gamma_{\rm 3B}=-\frac1N\frac{dN}{dt}=-2\ \text{Im} H\approx2 \eta \frac{\partial H}{\partial \theta}=2\eta C_\theta.
\end{align}
The extra parameter $\eta$ should be positive which is the situation that in the hyper-spherical formalism, the incoming wave of three fermions has larger magnitude than the outgoing wave due to the loss occurring at short distances. The exact value of $\eta$ depends on the details of three-body loss channels.

All the above universal relations regarding the three-body contact $C_\theta$ can be derived via an effective field theory approach. 
We model the system by the Lagrangian density
\begin{align}
\mathcal{L}=&\psi^{\dagger}\left(i\partial_{t}+\dfrac{\nabla^2}{2}\right)\psi+\sum_{a=x,y} d_{a}^{\dagger}\left(i\partial_{t}+\dfrac{\nabla^2}{4}-\nu_{0}\right)d_{a}\notag \\
&-\dfrac{g_{0}}{4}\sum_{a=x,y}\left\{d_{a}^{\dagger}[\psi(i\partial _{a}\psi)-(i\partial _{a}\psi) \psi]+h.c.\right\}\notag \\ 
&-v_3\psi^\dagger \left(\sum_{a=x,y} d^\dagger_ad_a\right)\psi\label{L},
\end{align}
which is assumed applicable up to the momentum cutoff $\Lambda$.
Here $\psi$ are the field operators for the fermions, $d_a$ are the dimer field operators for the close channel molecules responsible for the $p$-wave Feshbach resonance; the index $a$ takes $x$ or $y$ as the $p$-waves are two-fold degenerate in two dimensions.  The model (\ref{L}) has been employed to derive the existence of three-body super Efimov bound states at the $p$-wave interaction resonance through a renormalization group calculation \cite{Nishida}. 

To identify the operator for $C_\theta$ within our model (\ref{L}) to carry out further calculations, we first need to renormalize the bare coupling constants $\nu_0$, $g_0$ and $v_3$ in terms of $s$, $R$ and $\theta$. In the two-body sector, by solving the Lippmann--Schwinger equation for two scattering fermions with relative energy $E=k^2$ and matching the resultant $T$-matrix for the scattering with the phase shift $\cot\delta(k)=-1/(sk^2)+(2/\pi) \ln (kR)$, we obtain the renormalization relations
\begin{align}
&\frac{1}{s}=-\frac{16\nu _0}{g_0^2}+\frac{\Lambda^2}{\pi},\label{s}\\
&\frac{2}{\pi}\ln (R\Lambda)=-\frac{16}{g_0^2}.\label{r}
\end{align}
Across a $p$-wave Feshbach resonance in atomic gases, while $1/s$ flips its sign, usually the effective range $R$ remains positive and almost unchanged and shall be of order the van de Waals length $\ell_{\rm vdW}$. Equation (\ref{r}) indicates that the momentum cutoff $\Lambda$ must satisfy $\Lambda<1/R$; this condition is expected since the applicable energy regime of Eq.~(\ref{L}) shall be much smaller than the energy scale associated with the van de Waals length $\ell_{\rm vdW}$. We choose the cutoff to satisfy $\sqrt {E_ L}\ll\Lambda\ll R$, where $E_{ L}$ is of order the typical low energy scales of interest. Nevertheless, Eqs.~(\ref{s}) and (\ref{r}) show that a $p$-wave Feshbach resonance can be modeled by fine-tuning $\nu_0$.

Combined with Eqs.~(\ref{s}) and (\ref{r}), we can first use the adiabatic theorems (\ref{cs}) and (\ref{cr}) to identify the two-body contact operators as
\begin{align}
 C_s\equiv-\frac{\partial H}{\partial s^{-1}}=&\int d^2\mathbf r \left(\frac{g_0^2}{16}\sum_{a=x,y} d^\dagger_ad_a\right),\label{csf}\\
 C_R\equiv-\frac{\partial H}{\partial R^{-1}}=&\int d^2\mathbf r \left[\frac{g_0^2R}{8\pi}\sum_{a=x,y} d^\dagger_a\left(i\partial_t+\frac{\nabla^2}{4}\right)d_a \right]. \label{crf}
\end{align}
Within the model (\ref{L}), we also define the probability of finding two fermions weighted by their center of mass energy
\begin{align}
C_{s,P}\equiv\frac{g_0^2}{16}\sum_{a=x,y}\int d^2\mathbf r d^\dagger_a\left(-\frac{\nabla^2}{4}\right)d_a \label{csp},
\end{align}
which appears in Eq.~(\ref{t-b}).

\begin{figure}[t]
\includegraphics[width=3 in]{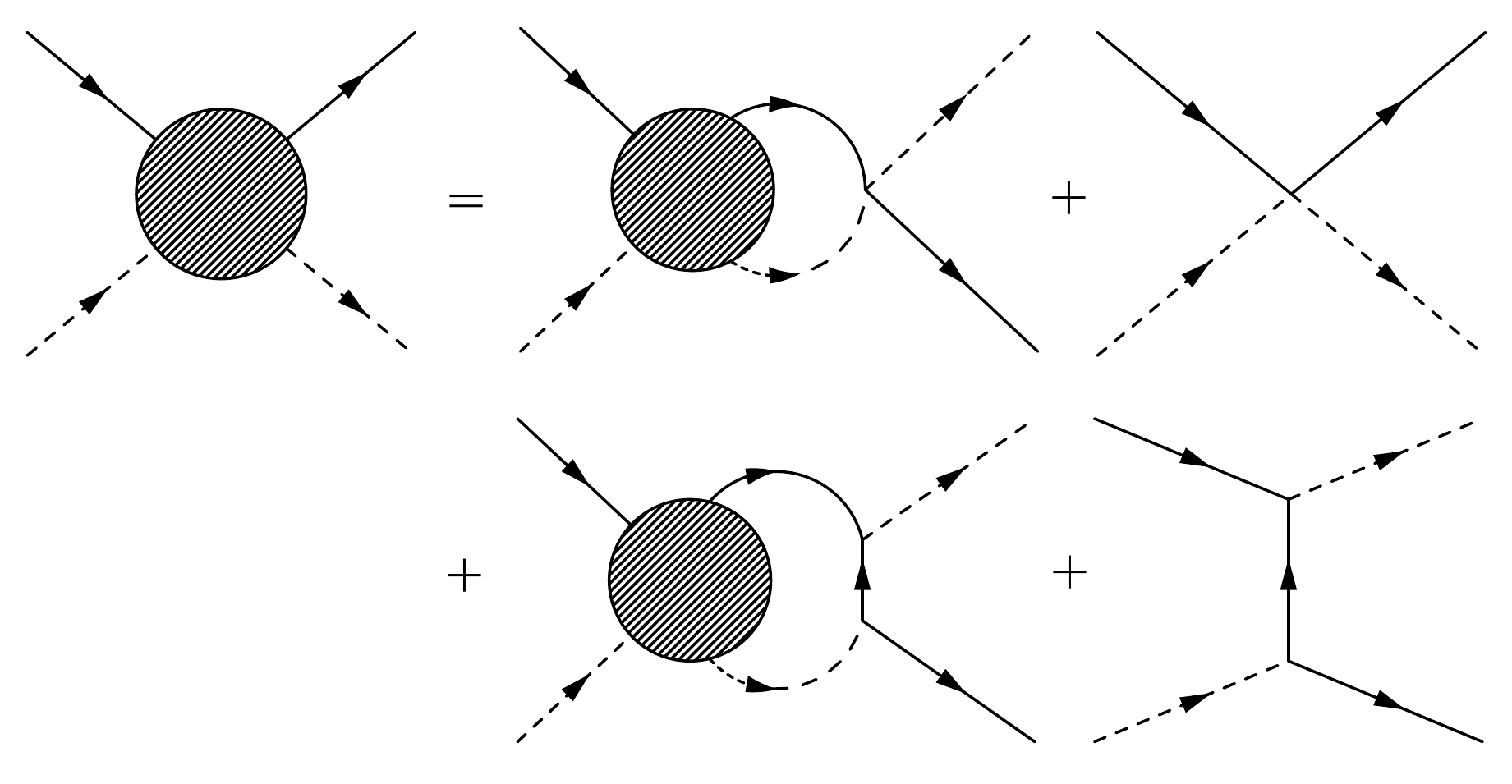}
\caption{Feynman diagrams for the ``three-body" atom-dimer scattering. The solid lines represent the bare fermion propagators and the dashed lines represent the full dimer propagators.}
\label{dia1}
\end{figure}

To renormalize $v_3$, 
we consider a ``three-body" scattering problem that in the center of mass frame a dimer with incoming momentum $\mathbf{p}$ in internal state $a$ scatters off a fermion and ends up with outgoing momentum $\mathbf{k}$ in internal state $b$. Instead of the Lippmann--Schwinger equation, the three-body scattering amplitude for the process, $B_{ab}(\mathbf{p},\mathbf{k},E)$ with relative energy $E$, satisfies the Skorniakov--Ter-Martirosian (STM) equation \cite{stm,review}. According to the diagrams for the scattering processes given in Fig.~\ref{dia1}, at a $p$-wave resonance $1/s=0$, the 
STM equation is
\begin{align}
 B_{ab}(\mathbf{p},\mathbf{k},E)=& K_{ab}(\mathbf{p},\mathbf{k},E)+\sum_{c=x,y}\int\frac{d^2q}{(2\pi)^2}B_{ac}(\mathbf{p},\mathbf{q},E)\notag\\
&\times \frac{8\pi K_{cb}(\mathbf{q},\mathbf{k},E)}{(E-3q^2/4)\ln(\sqrt{-E+3q^2/4}R)},\label{stm}
\end{align}
with
\begin{align}
K_{ab}(\mathbf{p},\mathbf{k},E)=&-\frac{v_3}{g_0^2}\delta_{ab}+\frac{(\mathbf{p+2k})_a\mathbf{(2p+k)}_{b}}{4(E-k^2-p^2-\mathbf{p}\cdot\mathbf{k})}.
\end{align}
The integral in Eq.~(\ref{stm}) is cutoff at $q=\Lambda$. 

We use Eq.~(\ref{stm}) to solve for the super Efimov bound states. 
By requiring the $\Lambda$-dependent divergence in the integral equation (\ref{stm}) countered by $v_3$, we find that the renormalized three-body coupling constant $\theta$ is given by
\begin{align}
\frac{v_3}{2\pi}\ln(\Lambda R)=1-\cot\left\{\frac{4}{3}\left(\ln[-\ln(\sqrt{3}\Lambda R/2)]-\theta\right)\right\}.\label{re}
\end{align}  
Equation (\ref{re}) agrees with the result from a perturbative renormalization group calculation in Ref.~\cite{Nishida}. 
Numerically, we have also confirmed that the super Efimov state binding energies $E_n$ indeed obey $\ln(-\ln|E_n|)\sim3\pi n/4 +\theta$ and do not depend on the cut-off $\Lambda$ when we use the renormalization relation (\ref{re}). 

By Eqs.~(\ref{ctheta}) and (\ref{re}), we identify the three-body contact operator
\begin{align}
C_\theta\equiv&\frac{\partial H}{\partial \theta}\nonumber\\
=& \frac{g_0^2}{3}\left[1+\left(1+4\frac{v_3}{g_0^2}\right)^2\right]\int d^2\mathbf r\psi^\dagger \left(\sum_{a=x,y} d^\dagger_ad_a\right)\psi.\label{cthetaf}
\end{align}
Based on Eq.~(\ref{cthetaf}), using the operator product expansion method in the three-body sector \cite{Braaten2011}, we derive the universal relations (\ref{i3}) and (\ref{n3}).

In summary, we have taken into account the newly discovered three-body super Efimov effect and introduced a new thermodynamic quantity, the three-body contact $C_\theta$, to two dimensional gases of identical fermions close to a $p$-wave interaction resonance. This contact reveals the three-body correlations of the many-body at short distances. We have shown how the contact affects physical observables such as the radio-frequency spectrum, the momentum distribution and the atom loss rate. Detection of the nonzero three-body contact $C_\theta$ via these observables would serve as a signature of the underlying super Efimov effect in the thermodynamic system. 

We thank Hui Zhai and Zheyu Shi for helpful input. This research was supported in part by NSFC Grant No. 11474179.

\end{document}